\documentclass[english,aps, preprint]{revtex4}
\usepackage{graphicx}

\makeatletter

\usepackage{babel}
\makeatother
\begin{document}


\title{Valley blockade quantum switching in Silicon nanostructures}

\author{Enrico Prati}
\affiliation{Laboratorio Materiali e Dispositivi per la Microelettronica,
Consiglio Nazionale delle Ricerche - IMM, Via Olivetti 2, I-20041
Agrate Brianza, Italy}
\email{enrico.prati@cnr.it}

\begin{abstract}
In analogy to the Coulomb and the Pauli spin blockade, based on the electrostatic repulsion and the Pauli exclusion principle respectively, the concept of valley blockade in Silicon nanostructures is explored. The valley parity operator is defined. Valley blockade is determined by the parity conservation of valley composition eigenvectors in quantum transport. A Silicon quantum changeover switch based on a triple of donor quantum dots capable to separate electrons having opposite valley parity by virtue of the valley parity conservation is proposed. The quantum changeover switch represents a novel kind of hybrid quantum based classical logic device.
\end{abstract}

\newpage

\maketitle

\section{Introduction}

The  selection of electrons based on the valley parity index in Silicon nanostructures doped with individual atoms is discussed, towards the realization of quantum switches operated in the valley blockade regime.

According to the Moore's law, the scaling down of the size of semiconductor devices is close to the limits imposed by quantum mechanics. A groundbreaking step consists of the transition from classical devices to
atomic-scale quantum devices. There, the architecture has ground on few electron and/or few dopant quantum dots. \cite{Loss98,DiVincenzo00} Deterministically implanted atoms immersed in the Si \cite{Shinada05} may be used to either store or manipulate information \cite{Prati08,Sanquer09}. A novel kind of hybrid quantum based classical logic devices will represent the natural evolution of mid-term applications towards atomic scale electronics and solid state qubit, which represent the ultimate application of group IV-based nanostructures. Here I define the valley blockade regime in single donor quantum dots and I suggest a new family of quantum devices for classical logic.


Major importance has been attributed to electron spin physics in silicon nanostructures and to its predominant role when looking for a workable Hilbert space for physical qubits. Valley states can compete with spin states to build a base of the Hilbert space of quantum devices. Such competition is generally considered an issue.\cite{Friesen03}
On the contrary, valley eigenstates provide a powerful quantum index themselves, as the valley index forms
a good quantum number.\cite{Rogge10} Here I suggest that the valley parity index allows to fully exploit the quantum properties of electrons in silicon nanostructures at cryogenic temperature.
Electronic states in multi-valley Silicon quantum dots have been extensively discussed after Hada and Eto. \cite{Hada03} Valley splitting in silicon nanostructures has been studied in both top gated Si/SiGe heterostructures \cite{Goswami06,Frucci10} and quantum dots \cite{Friesen03,Friesen06}, and in single donor quantum dots \cite{Rogge08,Prati08,Rogge10,Tan10}. The concept of valley blockade has been anticipated in the conclusion of Ref. \cite{Wunsch09} as possible alternative to spin blockade.  
The paper summarizes the key concepts of valley splitting in Silicon nanostructures and introduces the valley parity operator and the valley blockade regime. Next, a quantum changeover switch based on the valley parity conservation is discussed as an example of a new generation of quantum devices totally or partially based on the valley blockade regime.
The paper is organized as follows: in section II, the valley degeneracy in Silicon nanostructures is summarized. In section III, stability diagrams constrained of valley parity conserved transport through a single donor quantum dot are discussed. Section IV describes the conceptual and technological aspects of a quantum changeover switch based on the valley parity selection of electrons.

\section{Valley degeneracy breaking in Silicon nanostructures}

Unstrained bulk silicon is known for being an indirect bandgap semiconductor. When conduction electrons are confined close to an interface their six-fold valley degeneracy typical of bulk is splitted in a two-fold degeneracy plus a four-fold degeneracy associated to the symmetry breaking. Only the lowest valley doublet participates to transport at sufficiently low temperature. Unlike III-V compound semiconductors, the valley degeneracy is further lifted in single electron and single donor quantum dots close to the interface. This section reviews the valley splitting of quantum dots and donor quantum dots and introduces the definition of valley parity operator.

In the bulk Si effective mass theory, the wavefunction of 
conduction electrons is expressed by the
sum of contributions from the six degenerate valleys. In nanostructures, the symmetry breaking caused by the interface surfaces (both between Si and SiO$_2$ or Si$_{1-x}$Ge$_x$ and Si in quantum well) generates a tensile
strain in the Si. If the surface is perpendicular to the z axis, the degeneracy is lifted and the
two z valleys are lowered in energy so they are the only ones which play a role at sufficiently low
temperature. Effective mass theory correctly captures the essence of both the long wavelength and the atomic scale features. \cite{Friesen07} The Hamiltonian of the system in the single electron picture is:
\begin{equation}
H=H_0+V_v(z)=T(z)+V_{QW}(z)+V_{\phi}(z)+V_v(z)
\end{equation}
where
\begin{equation}
T=-\frac{\hbar}{2}\frac{\partial}{\partial z} \left( \frac{1}{m_l}\frac{\partial}{\partial z} \right)
\end{equation}
while $V_{QW}$ is the quantum well band offset along the z direction and $V_{\phi}(z)\approx -eEz$ is the potential generated by an electric field due for instance to a gate electrod. $V_v(z)=v_v \delta(z-z_i)$ is the valley coupling generated by an interface at the height $z_i$ and it is treated as a perturbation.
Said $F_{\pm z} (\textbf{r})$ the envelope functions, the electron states 
\begin{equation}
\psi(\textbf{r})=\Sigma_{j=\pm z} \alpha_j e^{ik_j z} u_{\textbf{k}_{j}}(\textbf{r}) F_{j} (\textbf{r})
\end{equation}
are characterized to leading order by their valley composition vector:
\begin{equation}
\alpha=(\alpha_{-z},\alpha_{+z})
\end{equation}
At the first order of approximation, the degenerate eigenvalues $\epsilon_0$ of the unperturbed hamiltonian are associated to the even and the odd combination of the two eigenstates and become:
\begin{equation}
\epsilon_{\pm}=\epsilon_0 + \Delta_0 \pm \left| \Delta_1 \right|
\end{equation}
where, said $F^{0}$ the first order approximation of the envelope function,
\begin{equation}
\Delta_0 = \int V_v (z) \left| F^{0} (z) \right|^2 dz
\end{equation}
and
\begin{equation}
\Delta_1 = \int e^{-2ik_0 z} V_v (z) \left| F^{0} (z) \right|^2 dz
\end{equation}
The $\epsilon_{\pm}$ correspond to the even (+ , excited) and odd (-, ground) combinations of the valley composition eigenvectors
\begin{equation}
\alpha_{\pm}=\frac{1}{\sqrt{2}}(e^{i \theta},\pm e^{-i \theta})
\end{equation}
respectively, where 
$e^{i \theta}=\frac{\Delta_1}{\left| \Delta_1 \right|}$
To simplify the notation, we define the valley parity operator 
\begin{equation}
K_v=\sigma_x K_0 
\end{equation}
where $K_0$ is the complex conjugation operator. The valley composition eigenvectors are eigenvectors of the valley parity operator as 
\begin{equation}
K_v \alpha_{\pm} = \pm \alpha_{\pm}
\end{equation}
The positive and negative eigenvalues $k_v$ may be expressed by a parity index $v$ so that $k_v=(-1)^v$ where v is either an even ($e$) or an odd ($o$) integer respectively. 
The valley parity index $v=e,o$ determines the eigenvalue $k_v$ of the valley parity operator $K_v$ on the valley composition eigenvector, which assumes the values $\pm 1$.
For a square well potential, the ground state alternates between valley parity $o$ and $e$ as a function of the width $L$. Typical valley splitting of Silicon quantum dots and 2-dimensional electron systems (2DES) are of the order of few hundreds of meV. Higher values are reached in donor quantum dots, in the range from 1.5 meV to tens of meV.
Donor quantum dots are obtained by either random diffusion or deterministic doping \cite{Shinada05} in the channel of a Silicon device. The state of a donor close to the oxide interface in the Silicon channel is hybridized with the quantum dot state formed at the interface \cite{Rogge08}.
According to Ref. \cite{Rogge10}, the conservation of valley index is predicted for symmetric systems and for the large 2D confinement provided by the electric field, which suggests that the ground and first excited states, GS and ES, consist of linear combinations of the $\textbf{k}=(0, 0, k_z )$ valleys (with z in the electric field direction). As the momentum perpendicular to the tunneling direction is conserved, the valley parity index is also conserved in tunneling. 
The valley parity index becomes a good quantum number in competition with spin component projection along a direction. In the limit of strong coupling, the SU(4) symmetry may be achieved so valley and spin degrees of freedom are fully mixed. Eigenstates of opposite spin projection are separated in energy by the Zeeman spltting $\Delta E_z$, obtained by applying a static magnetic field of few Tesla. For a $g=2$ spin 1/2 particle, the Zeeman separation corresponding to a coupling electromagnetic field at a frequency of 40 GHz (160 $\mu$eV) is generated at about 1.43 T. Such splitting is enough to ignore the effects of temperature excitations at 300 mK, which corresponds to an energy of about 25 $\mu$eV. In Figure 1 three possible cases are considered. In the case $a$, the Zeeman splitting $\Delta E_z$ is much smaller than the valley splitting $\Delta E_v$. This case is relevant for spin based qubits. The case $b$ corresponds to similar Zeeman and valley splittings, so the Hamiltonian is not diagonal on the eigenfunctions of the spin projection along the magnetic field. The case $c$ represents a valley splitting much lower than the Zeeman splitting, which can be obtained by applying an intense magnetic field. The case $c$ creates simple conditions to describe the effects of the only valley parity index, like in the next section. 

\section{Valley blockade diagrams}

The stability diagram of a symmetric dot (same electron population in the leads) is  symmetric with respect to the $V_{ds}=0$ axis, upon a suitable normalization of the drain, source and gate capacitances respectively, so $C_d=C_s+C_g$. In this paragraph the asymmetric case of a quantum dot with two opposite valley parity index leads is examined. The asymmetry can be physically realized for instance by using leads made by two dimensional electron systems of different width $L_s$ and $L_d$ respectively, so the valley GS has even or odd parity. Alternatively, one may consider a symmetric device where the leads are identical and a miscut below one of the two barriers produces an exact valley parity flip across the barrier. 
As a result, the stability diagram is not symmetric. The stability diagram studied in this paragraph limits to the $N=0\rightarrow 1$ CB region. For the sake of simplicity, the spin of the electron is ignored.
The effect of the $e$ and $o$ leads is represented graphically by assigning a different color to the left (red, $o$ parity) Fermi energy $E_L$, and to right (blue, $e$ parity) Fermi energy $E_R$ (Figure 2 \textbf{a}-\textbf{d}). 
In Figure 2 the expected energy diagram of an ideal asymmetric Silicon quantum dot is represented for three interesting cases. In the case 1 and 2, the gate voltage is sweeped as a function of time from low to high values, at fixed $V_{ds}$. Consequently, once the $V_{ds}$ is fixed, the quantum dot is initially empty ($N=0$). The case 3 refers to the opposite direction of the sweep, so the ground state of the quantum dot is initially full ($N=1$, energy of the electron below the Fermi energy of both the leads). The case 1 and 2 differ for to the assumption about the selection rules on the relaxation of the electron from the ES to the GS. In the case 1, the transition ES$\rightarrow$GS in the quantum dot is strictly forbidden. In the case 2 the relaxation is allowed, so once the current is turned on, the quantum dot becomes full ($N=1$) much faster than the scanning time scale, so the Coulomb repulsion blocks the current. The (donor) quantum dot is characterized by an odd ground state (red) and an even excited state (blue) so that $E_{ES}=E_{GS}+\Delta$. 
In the cases 1-3, the upper part of the stability diagram is generated by the processes \textbf{a} and \textbf{b} of Figure 2. Differently from standard Coulomb blockade diagrams of a two level system, the ES does not participate to tunneling because of its opposite valley index parity. 
In the lower part of the stability diagram, the selection of even (blue) electrons for tunneling generates a shifted triangle (processes \textbf{c} and \textbf{d}). In the case 1, the transition from ES to GS is forbidden so the current is allowed also for $eV_{sd}<\Delta$. In cases 2 and 3 one electron is present in the quantum dot for $eV_{sd}<\Delta$ because of the relaxation to the GS in the quantum dot or by hyphotesis (reverse sweep), so the Coulomb blockade stops the current, while at $eV_{sd}>\Delta$ both the two tunneling channels (\textit{o} and \textit{e}) are in the bias window.

\section{Valley parity quantum changeover switch}
As a possible application of valley blockade, a new family of classical logic devices based on pure quantum properties of electrons is proposed.
In the previous paragraph the valley parity of the incoming electrons is determined by the combination of the properties of the leads and the Si interface roughness at the barrier. In the following, the valley parity is selected at the source side by a donor quantum dot whose energy level alignement with the Fermi energy of the source $E_s$ is determined by a control gate. Valley parity flip is prevented by the ideality of the sample. Donor quantum dots proved to have a high valley splitting (1.5 meV to tens of meV) in Silicon nanostructures. Irrespectively from the spin, the electrons flowing through the GS and the ES have a $o$ and $e$ valley parity respectively.
The control gate switches the donor quantum dot from a 0 to a 1 electron state, in order to align alternatively the GS and the ES with the Fermi energy of the source. The source is supposed to contain spin and valley unpolarized electrons. The device is terminated by two drains $D_1$ and $D_2$ through two arms, each containing a donor quantum dot, with $E_s > E_{d1},E_{d2}$. Each donor quantum dot at the drain side is permanently tuned to a specific and opposite valley parity. Consequently, the current can be directed towards either the drain 1 or 2, according to the valley parity index of the electrons selected by the control gate.  
Figure 3 shows the 3D band diagram of the device. In the case \textbf{a} the control gate is tuned to select odd parity electrons, which are attracted by the drain 1. In the case \textbf{b} the control gate selects the even parity electrons, so the current flows to drain 2. Figure 4 shows the working principle of the device, which could be used to generate two opposite valley polarized currents by applying a high frequency square wave switching potential to the control gate. The valley parity of the electrons (odd in red, even in blue) is selected by virtue of a donor quantum dot on the source side, whose energy is tuned by a control gate. Each arm allows the tunneling of either odd and even parity electrons. The switching of the control gate allows the separation of the opposite index electrons in the two arms.
The discussed changeover switch implements classical logic by means of a pure quantum effect and can be further developed towards more complicated logic based on the valley parity polarized electrons.

\section{Conclusion}
In this paper, the physics of valley splitting in Silicon nanostructures is reviewed. The concept of valley blockade is explored, together with the valley parity operator. A changeover switch based on a triple of donor quantum dots is proposed as a possible application to switch a current and select electrons through the valley parity. It represents the ground to develop both classical logic devices based on quantum properties and quantum devices for qubits based on the valley parity composition eigenvectors. 

\begin{acknowledgments}
The author would like to thank Matteo Belli for the useful discussions and Diana Caputo for the graphical representation of the device in Figure 4. 
\end{acknowledgments}

\newpage

\section{Figures}

\begin{figure}[h]
\begin{center}
\includegraphics[width=0.46\textwidth]{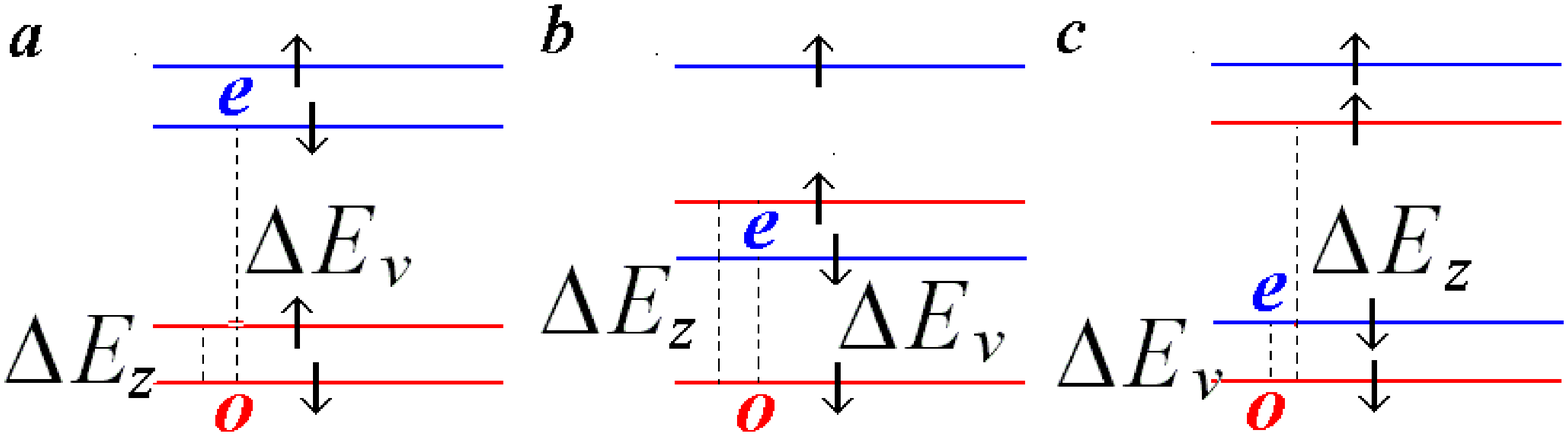}
\end{center}
\caption{\textbf{Valley blockade quantum switching in Silicon nanostructures - Enrico Prati}
\\The three possible configurations of Zeeman spin and valley splitting. In the case \textit{a}, the valley splitting is much higher than the Zeeman spin splitting generated by an external static magnetic field. This condition is particularly suitable for spin qubits. In the case \textit{b}, the two splittings are comparable. In the case $c$ the Zeeman spin splitting is much higher than the valley splitting.
}
\label{Figure1}
\end{figure}

\begin{figure}[t]
\begin{center}
\includegraphics[width=0.46\textwidth]{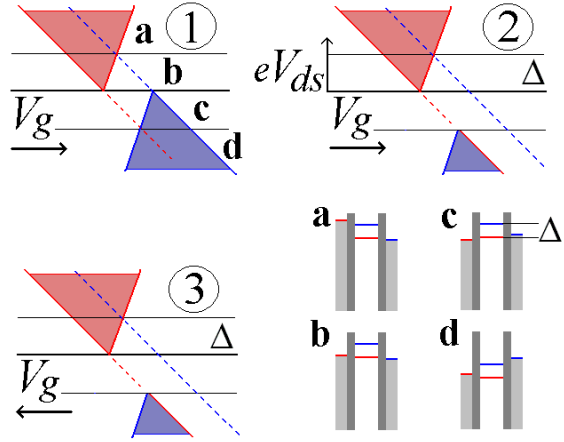}
\end{center}
\caption{\textbf{Valley blockade quantum switching in Silicon nanostructures - Enrico Prati}
\\Stability diagram in the case of opposite valley parity leads for forward (1 and 2) and reverse (3) gate voltage $V_g$ scans. In the case 1, the transition from the excited to the ground state in the quantum dot is strictly forbidden. The valley blockade implies two shifted current diamonds on the first peak, originated by the four possible situations shown in the configurations \textbf{a}-\textbf{d}. The case 2 consists of the previous one with the further hyphotesis that the process $ES\rightarrow GS$ is allowed, so the Coulomb blockade occurs after the first electrons occupy the quantum dot during the sequential tunneling. In case 3, because of the reversed gate voltage sweep direction, the quantum dot is assumed to contain one electron in the GS when the sweeps start.}
\label{Figure2}
\end{figure}

\begin{figure}[t]
\begin{center}
\includegraphics[width=0.46\textwidth]{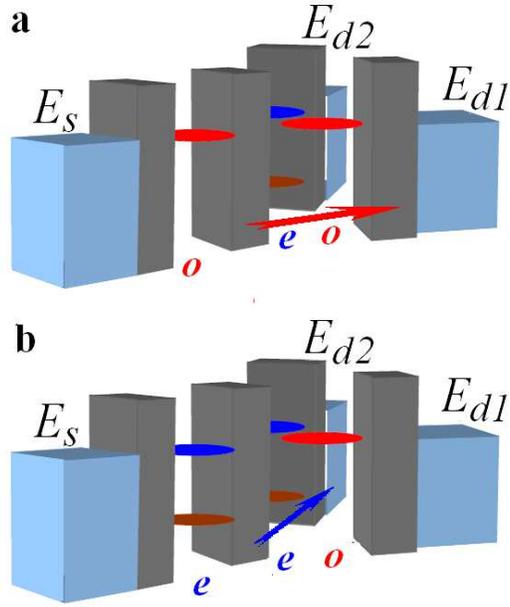}
\end{center}
\caption{\textbf{Valley blockade quantum switching in Silicon nanostructures - Enrico Prati}
\\
Energy diagram of the valley blockade quantum switch based on three donor quantum dots in the two possible configurations. In the case \textbf{a}, the first donor (on the left) - which is controlled by a control gate - selects the odd electrons for tunneling. Conversely, in the case \textbf{b}, it selects the even electrons. The two gates $V_{g1}$ and $V_{g2}$ of the arms 1 and 2 of the device repectively, are tuned in order to align the energy levels of the donor quantum dots in the arms in a $e$ and $o$ states. The flow is directed to one of the two arms by selecting the valley parity index in the first donor at the source side.}
\label{Figure3}
\end{figure}

\begin{figure}[t]
\begin{center}
\includegraphics[width=0.46\textwidth]{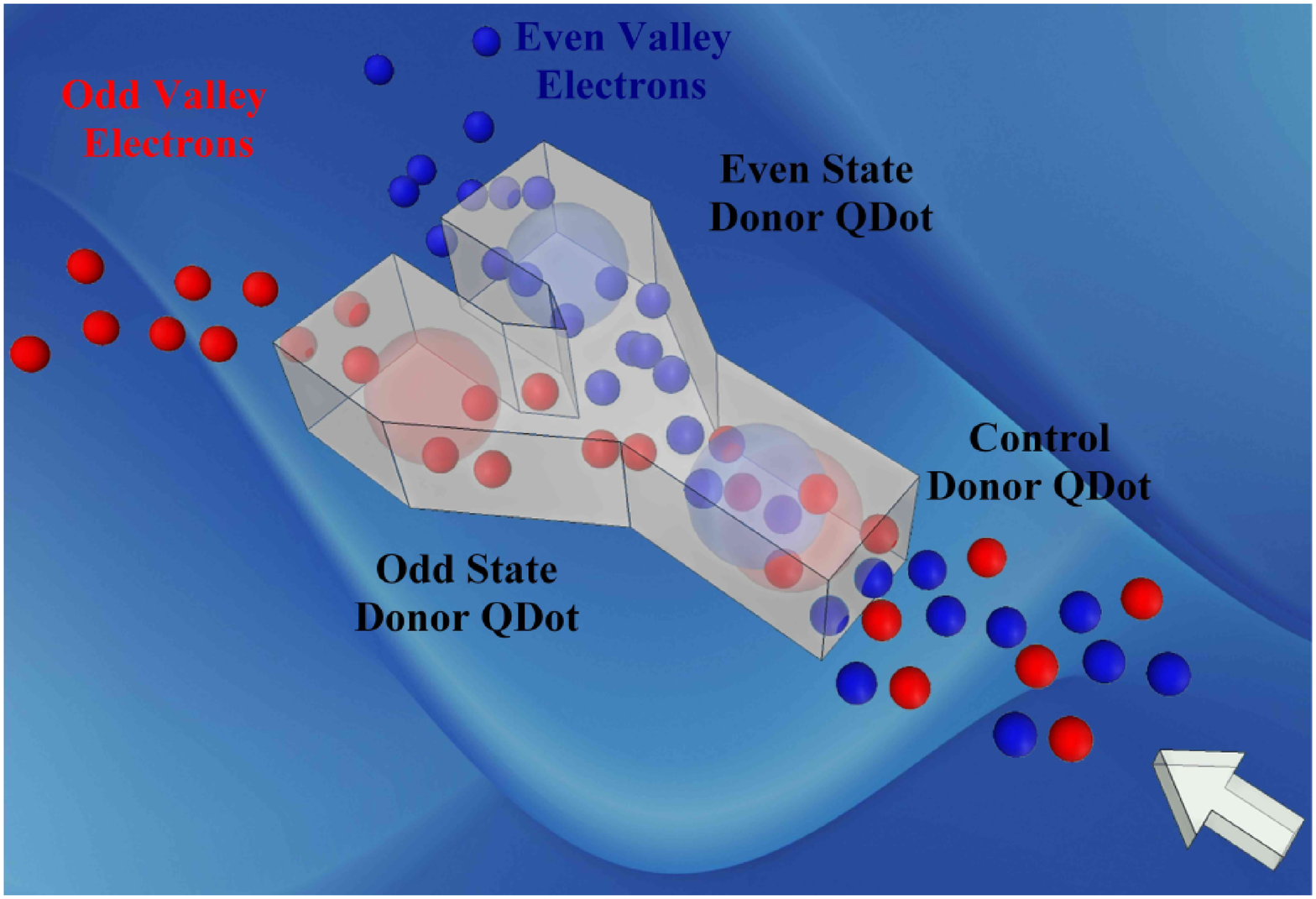}
\end{center}
\caption{\textbf{Valley blockade quantum switching in Silicon nanostructures - Enrico Prati}
\\
Graphical representation of the concept of the valley changeover switch. The electrons are selected by a gated donor quantum dot which can be tuned alternatively in the odd (ground, red) parity state, and in the even (excited, blue) parity state. The channel splits in two arms containing a donor quantum dot each, tuned respectively on the odd and the even state. According to the valley parity sign, the electrons alternatively flow either to the left or to the right drains.}
\label{Figure4}
\end{figure}


\begin{thebibliography}{10}
\bibitem{Loss98} D. Loss and D. P. DiVincenzo: Phys. Rev. A \textbf{57} (1998) 120
\bibitem{DiVincenzo00} D. P. DiVincenzo, D. Bacon, J. Kempe, G. Burkard and K. B. Whaley: Nature \textbf{408} (2000) 339
\bibitem{Shinada05} T. Shinada, S. Okamoto, T. Kobayashi and I. Ohdomari: Nature 437, 1128-1131 (2005)
\bibitem{Sanquer09} M. Pierre, R. Wacquez, X. Jehl, M. Sanquer, M. Vinet and O. Cueto: Nature Nanotechnology \textbf{5} (2010) 133
\bibitem{Prati08} E. Prati, R. Latempa and M. Fanciulli: Phys. Rev. B \textbf{80} (2009) 165331
\bibitem{Friesen03} M. Friesen, P. Rugheimer, D. E. Savage, M. G. Lagally, D. W. van der Weide, R. Joynt, and M. A. Eriksson, Phys. Rev. \textbf{B 67}, 121301(R) (2003)
\bibitem{Rogge10} G. P. Lansbergen, G. C. Tettamanzi, J. Verduijn, N. Collaert, S. Biesemans, M. Blaauboer and S. Rogge: Nano Letters \textbf{10} (2010) 455
\bibitem{Hada03} Y. Hada and M. Eto, Phys. Rev. \textbf{B 68}, 155322 (2003)
\bibitem{Goswami06} S. Goswami, K. A. Slinker, M. Friesen, L. M. McGuire, J. L. Truitt,
C. Tahan, L. J. Klein, J. O. Chu, P. M. Mooney, D. W. Van Dreweide, R. Joynt,
S. N. Coppersmith and M. A. Eriksson: Nat. Phys. \textbf{3}, 41 (2006)
\bibitem{Frucci10} G. Frucci, L. di Gaspare, F. Evangelisti, E Giovine, A. Notarigiacomo, V. Piazza, and F. Beltram, Phys. Rev. \textbf{B 81}, 195311 (2010)
\bibitem{Friesen06} M. Friesen, Appl. Phys. Lett., 89, 202106 (2006)
\bibitem{Rogge08} G. P. Lansbergen, R. Rahman, C. J. Wellard, I. Woo, J. Caro, N. Collaert, S. Biesemans, G. Klimeck, L. C. L. Hollenberg and S. Rogge: Nature Physics \textbf{4} (2008) 656 
\bibitem{Tan10} K. Y. Tan, K. W. Chan, M. M\"ott\"onen, A. Morello, C. Yang, J. van Donkelaar, A. Alves, J.-M. Pirkkalainen, D. N. Jamieson, R. G. Clark and A. S. Dzurak: Nano Letters \textbf{10} (2010) 11
\bibitem{Wunsch09} B. Wunsch: Phys. Rev. B \textbf{79}(2009) 235408
\bibitem{Weitz96} P. Weitz et al., Surf. Sci. 361-362, 542-546 (1996)
\bibitem{Friesen07} M. Friesen, S. Chutia, C. Tahan, and S. N. Coppersmith, Phys. Rev. B \textbf{75}, 115318 (2007) 
\end{thebibliography}
\end{document}